\documentclass[aps,prd,twocolumn,10pt]{revtex4-2}

\usepackage[english]{babel}
\usepackage{indentfirst}
\usepackage[latin1,utf8]{inputenc}
\usepackage{lmodern}
\usepackage[T1]{fontenc}
\usepackage[pdftex]{graphicx}
\usepackage{setspace}
\usepackage{hyperref}
\usepackage{color}

\usepackage{cellspace}
\usepackage{amsmath,amsfonts,amssymb}

\begin{document}

\title{Dispersion and focusing of cosmic rays in magnetospheres}

 \author{J.~Hirtz}
 \author{I.~Leya}
 \affiliation{Space Research and Planetary Sciences, Physics Institute, University of Bern, Sidlerstrasse 5, 3012 Bern, Switzerland}

\begin{abstract}
Simulating the irradiation of planetary atmospheres by cosmic ray particles requires, among others, the ability to understand and to quantify the interactions of charged particles with planetary magnetic fields.
 Here we present a process that is very often ignored in such studies; the dispersion and focusing of cosmic ray trajectories in magnetospheres.
 The calculations were performed using our new code CosmicTransmutation, which has been developed to study cosmogenic nuclide production in meteoroids and planetary atmospheres and which includes the computation of the irradiation spectrum on top of the atmosphere.
 Here we discuss effects caused by dispersion and focusing of cosmic ray particle trajectories.
\end{abstract}

\maketitle

%
% ======================================================================================================
%
\section{Introduction}
\label{intro}

The interactions of cosmic ray particles, either galactic cosmic ray (GCR) particles or solar cosmic ray (SCR) particles, with matter produce a large variety of so-called cosmogenic nuclides.
 The range of possible applications is impressively large, including the study of meteorites and the lunar surface, where cosmogenic nuclides are used to constrain cosmic ray exposure histories, sizes, collision events, and terrestrial ages.
 In addition, studies of cosmogenic nuclides in terrestrial surface rocks are often used to constrain ages, glacial coverages, erosion rates, and uplift rates.
 Nuclide production in the terrestrial atmosphere is an useful tool for archaeology, hydrology, and oceanography.
 While this short list is already impressive, it is by far not complete.
 For recent reviews see, e.g., \cite{Dunai2010,DavidLeya2019}.
 
While understanding and quantifying cosmogenic nuclide production in meteorites and the lunar surface is relatively straightforward (for some critical aspects see, e.g., \cite{Ammonetal2008, LeyaMasarik2009, Lietal2017}), it is more difficult for the surface of Earth and its atmosphere due to the interactions with the Earth's magnetic field \cite{Argentoetal2015a, Argentoetal2015b, MasarikBeer1999}.

A major necessity for cosmogenic nuclide studies is the precise knowledge of the cosmic ray particle spectrum.
 However, experimental measurements are very limited.
 Hadron fluxes below 200~GeV have been measured above the Antarctic by the BESS experiment \cite{bess}, which is a balloon-borne set-up.
 In addition, there have been data from the AMS \cite{ams} and PAMELA \cite{pamela} collaborations using detectors orbiting Earth on the International Space Station and the Russian Resurs-DK1 spacecraft, respectively.
 The thus determined particle spectra are for a given location.
 For example, the BESS data are only valid for the flight trajectory of the balloon, which was in the latitude range between $60^\circ$ and $80^\circ$ and in the range of zenith angles between $0^\circ$ and $35^\circ$ (all in the coordinate system of the magnetic field).
 From such local data, the particle spectra for all other locations must be calculated using sophisticated models.
 Thus, the model calculations must, {\em first}, quantify the effect of the magnetic field on the galactic particle spectrum at the measured location to calculate the (original) cosmic ray spectrum without magnetic interactions, i.e., the spectrum outside the magnetosphere.
 In a {\em second} step, the magnetic interactions must be understood and quantified to calculate galactic particle spectra at any point of interest inside the heliosphere. 
 
The Earth's magnetic field shields us almost completely from solar wind and SCR particles and partly from GCR particles.
 The minimum energy needed for a GCR particle to cross the Earth magnetic field and to reach the top of the atmosphere decreases with geomagnetic latitude, resulting in a cosmic ray flux that is minimal at the equator and is increasing towards the poles.
 In practice, the minimum energy also depends on the direction of incidence and the particle type.
 For studying particle trajectories in magnetic fields, rigidities are better units than energies because particles with the same rigidity follow the same trajectory, independent of the particle type.
 The rigidity $R$ is given by:

\begin{equation}
R = \frac{p}{q} = \rho B_\bot %\color{red} = \frac{\sqrt{E_{Kin} (E_{Kin} + 2E_{0})}}{q~c}
\end{equation}

with $p$ the particle momentum, $q$ the particle charge, $\rho$ the gyroradius of the particle, and $B_\bot$ the intensity of the orthogonal magnetic field.

As a first (relatively crude) approximation, the Earth geomagnetic field can be seen as a simple dipole field centered in the Earth with an axis $\sim$11$^\circ$ tilted relative to the planetary spin axis.
 Note that this dipole approximation is far too simple because the interaction of the Earth magnetic field with the solar wind results in various significant modifications of the Earth magnetic field.
 For example, there is a compression of the magnetic field lines on the day-side and a stretching of the magnetic field lines on the night side.
 Consequently, the magnetic field that should be used in the calculations is not the Earth's internal magnetic field but a modified version of it, usually called external geomagnetic field or magnetospheric field.
 For more details on magnetospheric and atmospheric effects see, e.g., \cite{Schereretal2006}.

Approximating the geomagnetic field by a simple geocentric dipole, we can calculate the effective cut-off rigidity using the St\"ormer cut-off formula:

\begin{equation}
 R_{c} = \frac{M \times cos^{4} (\lambda) }{r^{2} (1 + (1 - cos^{3} (\lambda) \times cos (\epsilon) \times sin (\eta))^{1/2})^{2}}
\end{equation}
 where $M$ is the dipole moment, $r$ is the distance from the dipole center, $\lambda$ is the geomagnetic latitude, $\epsilon$ is the azimuthal angle, and $\eta$ is the angle from the local magnetic zenith direction \cite{Stoermer1950, Cooketal1991}.
 This cut-off effect produces large variations in the cosmic ray particle spectrum as a function of latitude and longitude.
 For example, assuming a magnetic field strength of $B_0\simeq~30~\mu$T, i.e., comparable to the Earth geomagnetic field at the top of the atmosphere, most cosmic ray particles with energies below 20~GeV are dispelled near the magnetic equator.
 These energy range represents $\sim$99\% of the GCR particle spectrum (in particles, not in energy) and the entire SCR particle spectrum.
 This clearly demonstrates why it is so important to study cut-off effects in great detail.
 Unfortunately, the picture of a simple cut-off rigidity calculated via the St\"ormer cut-off formula is far too simple as discussed below.

For a more precise estimate of the cut-off rigidity, theoretical approaches using backward trajectory calculations are employed.
 The fundamental problem is that the integration of the equation of motion of a charged particle in a magnetic field has no solution in closed form.
 To circumvent the problem, the usual approach is to calculate many individual trajectories and to distinguish them into allowed and forbidden ones.
 Doing so and starting in reverse kinematics from the top of the atmosphere at the location of interest, various trajectories are calculated backwards in time considering only effects caused by the magnetic field.
 In this way, trajectories are tested for different impact parameters: the zenith and azimuth angles and the rigidity.
 If a tested trajectory connects the top of the atmosphere to a location outside the magnetosphere in reverse kinematics, the same trajectory can be followed by a cosmic ray particle having the same rigidity in reverse direction, i.e., in normal kinematics.
 This is the case of an {\em allowed} trajectory.
 In general, the trajectories for high-energetic particles experience only little magnetic bending before escaping the magnetosphere.
 With lower rigidities, i.e., lower energies and/or higher charge states, the particles will suffer more geomagnetic bending before escape and finally, for those below a certain rigidity, escape from the magnetosphere is no longer possible, defining so-called {\em forbidden} trajectories.
 In such calculations, there is a series of allowed and forbidden trajectory bands called the {\em cosmic ray penumbra}.
 An example is shown in \autoref{std}, where we plot the penumbra at $40^\circ$ latitude and $0^\circ$ longitude.
 The zenith and azimuth of the incoming particle trajectories are $45^\circ$ and $0^\circ$, respectively.
 The calculation is for a planetary radius of $R_0$~=~6448~km, i.e., for the top of Earth atmosphere.
 For the magnetic field we assume a dipole field of strength $B_0$ = 30.2 $\mu$T centred at the Earth core.
 This figure clearly demonstrates that the geomagnetic cut-off is not sharp but is defined by allowed and forbidden trajectory bands.
 For computational reasons, however, the penumbra is often approximated by an effective cut-off ($R_{c}$), which is an average between the lowest ($R_{l}$) and the highest ($R_{u}$) rigidity of the penumbra \cite{Smart2000}.
 In the calculations, the effective cut-off rigidity $R_{c}$ is then considered as a hard cut-off: all rigidities above $R_{c}$ are allowed and all rigidities below $R_{c}$ are forbidden.

Here we present a new approach for calculating allowed and forbidden trajectories by considering for the first time focusing and dispersion of particle trajectories in the magnetosphere (\autoref{newvsold}).
 After comparing our new approach to earlier approaches, we discuss in some detail the newly developed algorithm (\autoref{algo}).
 The results are analysed in \autoref{ana} and some possible consequences are discussed in \autoref{consequences}.
 Finally, we present some conclusions and perspectives in \autoref{conc}.

\section{New versus previous approaches}
\label{newvsold}

In this paper, the used coordinates refer to latitude and longitude in the magnetospheric reference frame, not in the geographic coordinate system.
 
\begin{figure*}
\begin{minipage}{2\columnwidth}
\begin{center}
\includegraphics[width=\columnwidth]{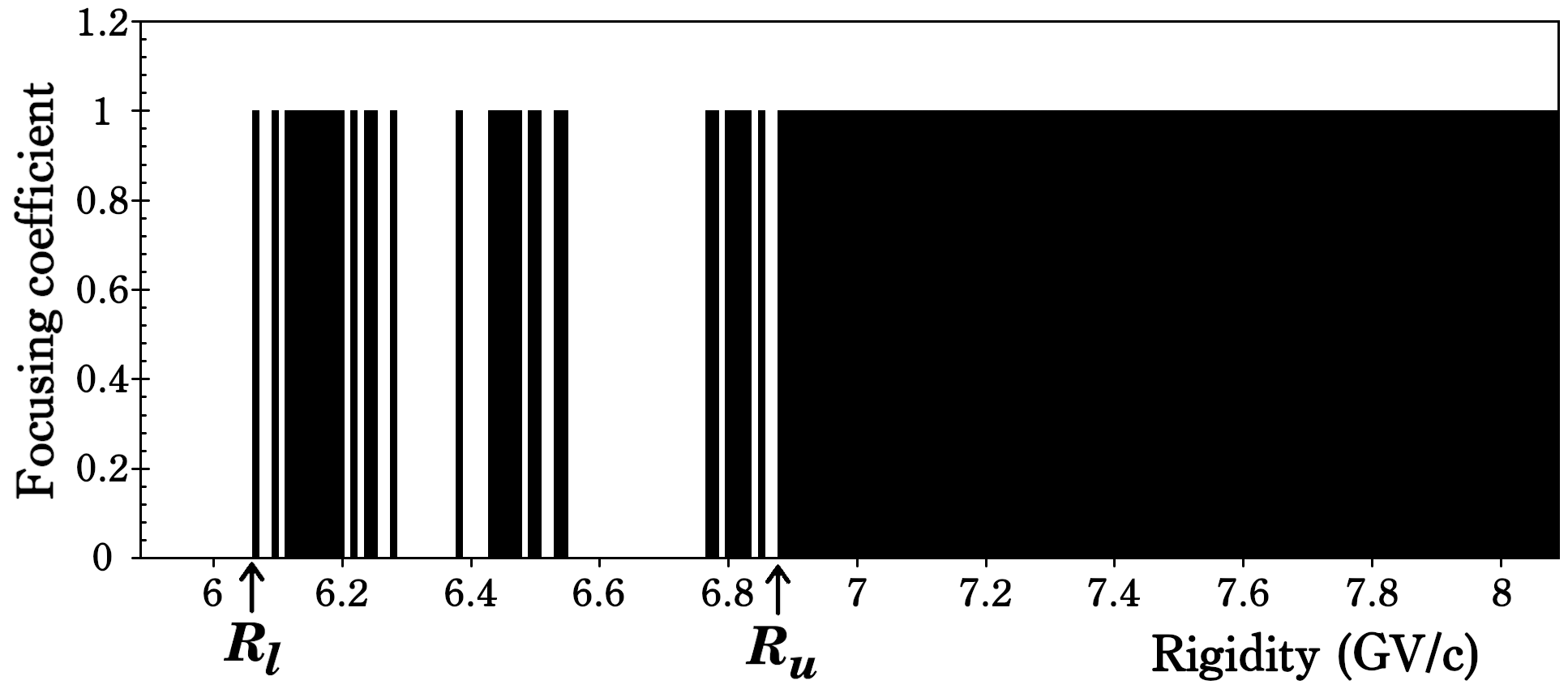}
\caption{\label{std} Example showing the penumbra, i.e., the region of allowed and forbidden trajectories calculated for the location: latitude~=~$40^\circ$ and longitude = $0^\circ$.
 The direction of the trajectory is zenith = $45^\circ$ and azimuth = $ 0^\circ$.
 The calculation is for a planetary radius of $R_0$~=~6448~km, i.e., for the top of the Earth atmosphere.
 The magnetic field is approximated as a dipole with field strength $B_0$ = 30.2 $\mu$T centred at the Earth core.
 Focusing and dispersion is not considered.}
\end{center}
\end{minipage}
\end{figure*}

 As mentioned above, the usual approach to calculate the cut-off effect is by using an inverse kinematics approach and testing for allowed and forbidden trajectories.
 For a given impact location (latitude and longitude) and direction (zenith and azimuth angles), the distribution of allowed and forbidden trajectories always follows the same structure \cite{Smart2000}, i.e., from high to low rigidities, there is the first forbidden trajectory at the so-called upper cut-off $R_u$.
 All trajectories above $R_u$ are allowed.
 Then there is the penumbra, in which allowed and forbidden trajectories alternate.
 The lower end of the penumbra is the so-called lower cut-off ($R_l$); below ($R_l$) all trajectories are forbidden.
 \autoref{std} depicts a typical example of a calculated penumbra.
 For practical applications, for each location of interest (latitude, longitude), such a penumbra calculation must be performed for all possible zenith and azimuth angles.

In the original approach, trajectories can only be allowed or forbidden.
 This is illustrated in \autoref{std}, where allowed trajectories are given the value 1 and forbidden trajectories are given the value 0.
 However, there are more physical effects by the Earth geomagnetic field than just simply allowing or forbidding trajectories.
 An interesting and relevant effect is that trajectories can be focused or dispersed.
 For example, protons with the same rigidity and the same direction of motion starting from a surface area of 1~m$^2$ at the top of the magnetosphere can reach the top of the atmosphere on a surface of 0.1~m$^2$, i.e., they are focused, or they can reach the top of the atmosphere on a surface area of 2~m$^2$, i.e., they are dispersed.
 For the following discussion we introduce the focusing coefficient, which we define as the ratio of the flux of cosmic ray particles having a certain set of impact parameters on top of the atmosphere relative to the flux of the very same particles outside the magnetosphere.
 With this definition, a focusing coefficient larger than $1$ indicates focusing, i.e., the flux density is (locally) increasing while the particles interact with the geomagnetic field.
 In contrast, a focusing coefficient lower than $1$ indicates dispersion, i.e., the flux density is (locally) decreasing due to the interactions with the geomagnetic field.
 In all previous calculations (e.g., \autoref{std}), allowed trajectories are given a focusing coefficient equal to $1$ (no focusing or dispersion).
 In our new approach, focusing coefficients in the range $]0,+\infty[$ correspond to allowed trajectories while focusing coefficients equal $0$ correspond to forbidden trajectories.
 In the following, the effect that caused focusing and dispersion of cosmic ray trajectories due to interactions with the geomagnetic field is called {\it funnel effect}.

\begin{figure*}
\begin{minipage}{2\columnwidth}
\begin{center}
\includegraphics[width=\columnwidth]{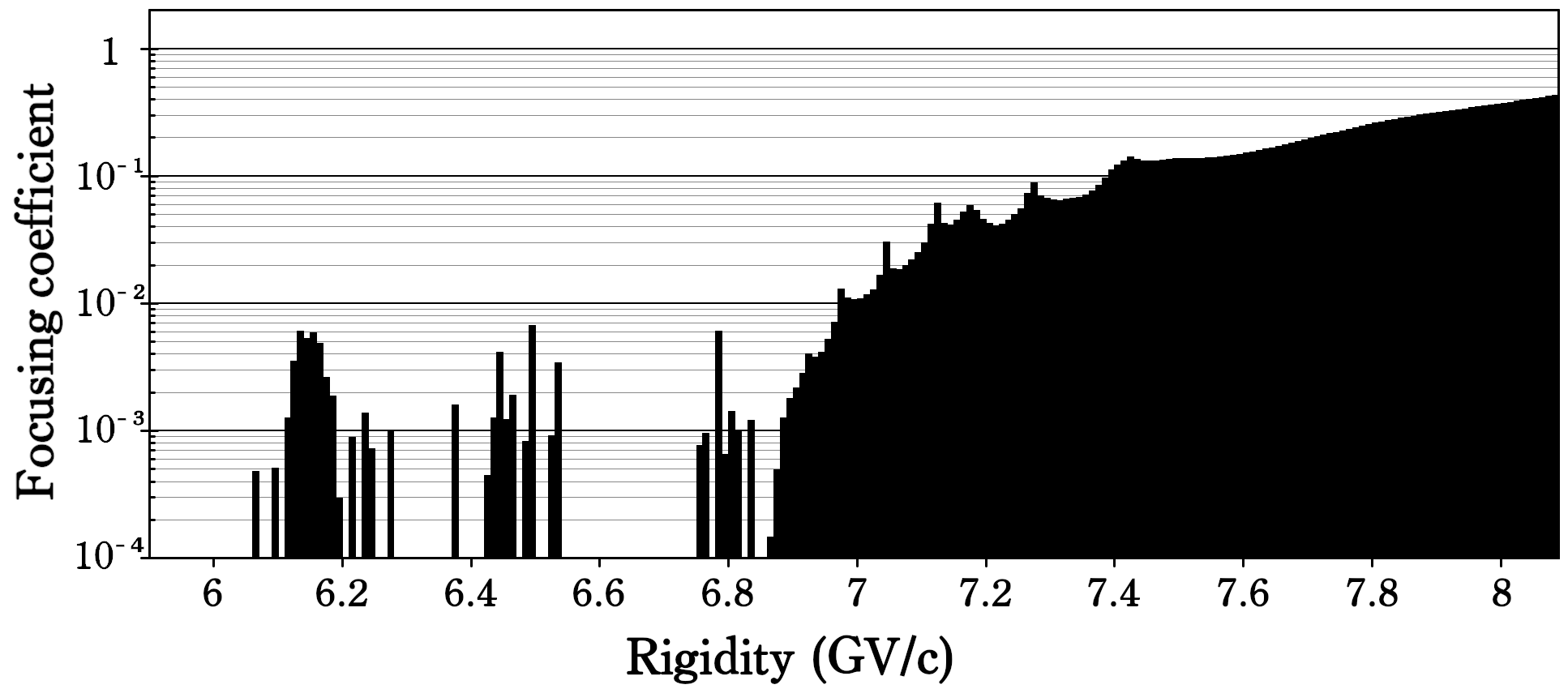}
\caption{\label{funnel} Same as \autoref{std} but this time considering the funnel effect, i.e., the focusing and dispersion of particle trajectories in the geomagnetic field.}
\end{center}
\end{minipage}
\end{figure*}

Considering the funnel effect adds an additional step to the standard algorithm computing the map of allowed and forbidden trajectories.
 This step includes testing whether a set of particle trajectories that are initially close in phase space, is focused, dispersed, or remains similarly close when travelling through the geomagnetic field.
 Note that interactions with the geomagnetic field are not changing particle rigidities.
 With this additional step of testing and with the definition of the focusing coefficient, the distribution of allowed and forbidden trajectories with values of only $0$ and $1$ (\autoref{std}) becomes a continuous distribution with values in the range [0,$+\infty$[ as shown exemplary in \autoref{funnel}.

Interestingly, a focusing effect of the magnetic field on the trajectories has been seen before \cite{Kasper1959}.
 This author found in his early numerical trajectory calculations that some particles with rigidities a factor of two higher than the cut-off rigidity that have been started outwards from the Earth with different azimuth and different zenith angles reached nevertheless a similar asymptotic direction far from the Earth (see also \cite{Smart2000}).
 However, to our knowledge, there have been no follow-on studies on this topic.
 
\section{algorithm}
\label{algo}

In all trajectory calculations, where there is a compromise between precision and CPU time, a trajectory calculated at a given rigidity and with a given direction of motion is assumed to be representative for a finite range of rigidities and directions, i.e., for a finite angular space.
 For example, for producing \autoref{std} and \autoref{funnel}, we used a rigidity binning of 10~MV/c and a zenith binning of 5$^\circ$.
 The binning of the azimuth depends on the zenith angle and is minimal at 12$^\circ$ for radial trajectories.
 An important question is, whether binning is a good approach for studying the chaotic behaviour of the penumbra.
 For a comprehensive discussion of the limitations and the accuracy of such calculations see \cite{Smart2000}.

The funnel effect for a given trajectory is computed by comparing this trajectory to other trajectories having very similar initial conditions when entering the magnetosphere.
 This step is done in direct kinematics.
 In practice, eight trajectories with slightly different initial positions but with the same initial direction as well as eight trajectories with slightly different initial directions but with the same initial position are tested.
 We then start the trajectory calculations of the 16 particles at a location far away from Earth where the magnetic field is weak.
 In practice we have chosen 9 times the Earth radius for this location.
 We then calculate the trajectories until the particles reach the top of the atmosphere.
 This calculation is done twice, once without any effects by the geomagnetic field and once by fully considering the interactions with the Earth magnetic field.

Here we give a very rough numerical example on how to understand focussing coefficients.
 Let us assume a set of eight particles all with the same initial position and the same rigidity.
 We assume for the eight particles slightly different travel directions, i.e., they form a cone.
 In our example the opening angle of the cone is $1^\circ$ and we start with a distance of 5000~km above the Earth atmosphere.
 Without magnetic field and without funneling, these particles cover, after a journey of 5000~km, an area of $\sim$5981~km$^2$.
 Adding now the interactions with the geomagnetic field, changes this area.
 The ratio of areas calculated without considering magnetic field interactions relative to the ones calculated by fully considering magnetic field interactions corresponds to the {\it focusing coefficient}.
 The new area covered by the set of particles can be larger than $\sim$5981~km$^2$, which indicates that fewer particles will reach the area.
 In this case, the ratio of areas calculated without magnetic field relative to the ones calculated with magnetic field, i.e., the focussing coefficient, is lower than $1$, indicating dispersion.
 The ratio can also be larger than $1$, indicating that trajectories are focused on a smaller area.

Calculating allowed and forbidden trajectories can be very CPU time consuming.
 The computational time depends on the precision wanted for the trajectories but also and predominately on the complexity of the Earth magnetic field used for the calculations \cite{Smart2000}.
 Trajectories are defined via three parameters (zenith, azimuth, and rigidity), which are all continuous.
 For practical applications, these parameters are binned, with all the problems and short-comings discussed earlier.
 Considering the funnel effect requires the computation of additional trajectories, which significantly increases CPU time consumption.
 To be more precise, the CPU time needed in our examples to calculate the funnel effect (\autoref{std} and \autoref{funnel}) was approximatively one order of magnitude higher than for the computation of the penumbra without considering the funnel effect.

One world of caution, the algorithm developed by us is likely biased.
 We checked the results for robustness by using slightly different versions, varying the initial phase space, and performing the same tests in reverse kinematics.
 We found that the focusing coefficient converges relatively quickly for the different tested versions with increasing energy, i.e., our approach can be considered robust at higher energies.
 We also observed, however, that our approach using direct kinematics to compute the funnel effect is more sensitive to dispersion of trajectories than to focusing, especially in the penumbra.

In Summary, there are two major differences between our new approach and earlier approaches.
 {\it First} and most importantly is the consideration of focusing and dispersion of trajectories, called funnelling.
 {\it Second}, all earlier approaches approximated the penumbra by a hard cut-off rigidity $R_c$ to save computer time.
 In contrast, our model considers the complete structure of the penumbra.

\section{analysis}
\label{ana}

Here we discuss some of the possible consequences of the funnel effect by simply comparing results obtained in calculations performed by considering the funnel effect to results obtained in the same type of calculations but without considering the funnel effect.
 Because of practical importance, we focus in this discussion on the effect of the funnel effect on the calculated GCR spectra on top of the Earth atmosphere.

By comparing the data shown in \autoref{std} that have been obtained without considering the funnel effect to the results given in \autoref{funnel}, which have been obtained by considering the funnel effect, it becomes clear that the latter calculations give much more complex results.
 Without funnel effect, trajectories are either allowed or forbidden, i.e., the focussing coefficient can only be zero or 1.
 This on-off-behaviour naturally results in significant non-physical discontinuities.
 The results obtained by considering the funnel effect are much more physical reasonable.
 The changes from allowed to forbidden trajectories occur at small focussing coefficients, producing less discontinuities.
 By comparing both figures, it is important to emphasize that trajectories that are allowed in the calculation without funnel effect are still allowed in the calculation with funnel effect and vice versa, i.e., trajectories that are forbidden without funnel effect are still forbidden with funnel effect.
  The major impact of the funnel effect is on the focusing coefficient and this effect can be very significant.
 For example, in the rigidity range of about 7~GV/c all trajectories are allowed in \autoref{std}.
 For the calculations with funnel effect, the trajectories are still allowed but the focussing coefficient is in the range $10^{-2}$, i.e., the cosmic ray flux is reduced by two orders of magnitude due to the dispersion of trajectories (\autoref{funnel}).
 This finding holds true for all tested and calculated penumbra regions.

\begin{figure*}
\begin{minipage}{2\columnwidth}
\begin{center}
\includegraphics[width=0.95\columnwidth]{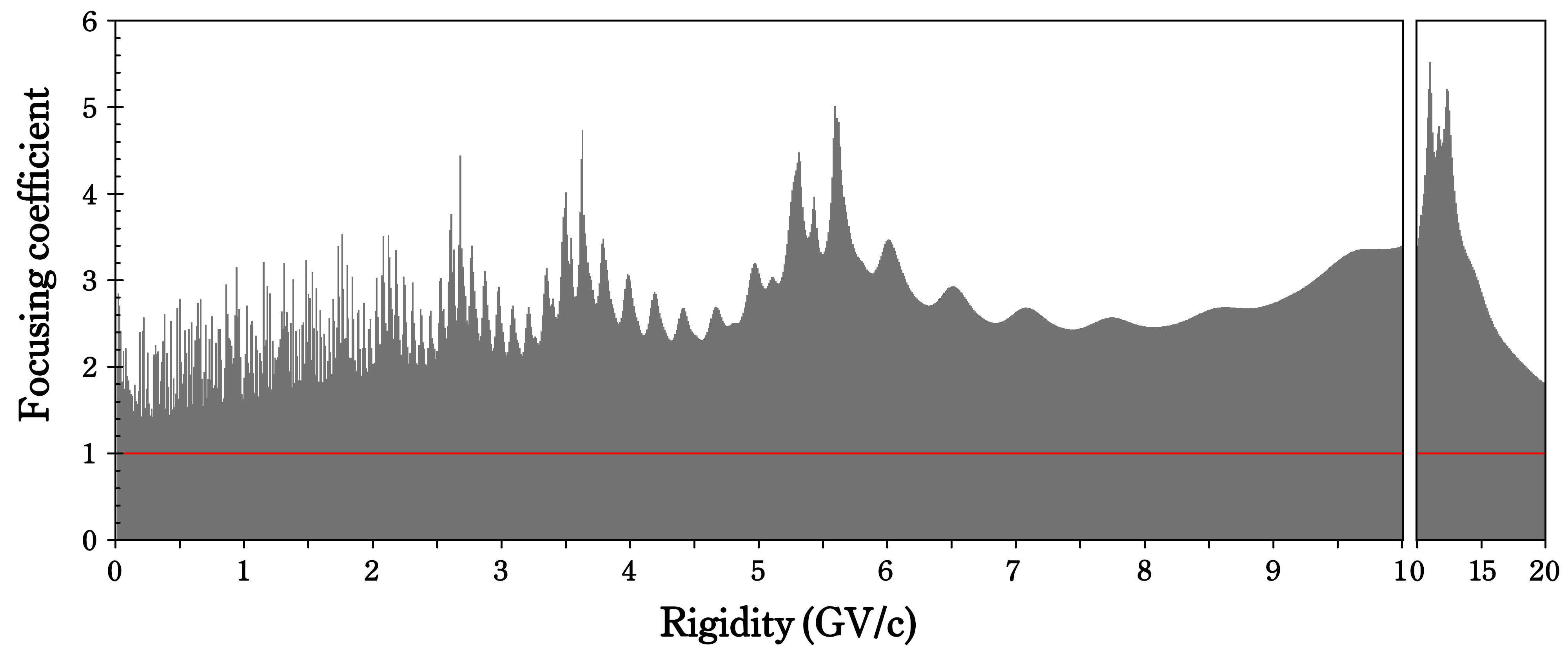}
\caption{\label{80} Example showing the penumbra for the location: latitude~=~$80^\circ$ and longitude = $0^\circ$.
 The direction of the trajectory is zenith = $45^\circ$ and azimuth = $ 0^\circ$.
 The calculation is for a planetary radius of $R_0$~=~6448~km, i.e., for the top of the Earth atmosphere.
 The magnetic field is approximated as a dipole with field strength $B_0$ = 30.2 $\mu$T centred at the Earth core.
 The funnel effect is considered.
 The red line, which is shown to guide the eyes, corresponds to a focusing coefficient of 1 (see text).}
\end{center}
\end{minipage}
\end{figure*}

\autoref{80} depicts the results of our trajectory calculations obtained by fully considering the funnel effect.
 The calculation is for a latitude of 80$^\circ$ and a longitude of 0$^\circ$.
 The direction of the trajectory in zenith angle is 45$^\circ$ and the azimuth is 0$^\circ$.
 The Earth magnetic field is again approximated as a dipole field with a strength of B$_0$ = 30.2 $\mu$T.
 At this latitude there is no real penumbra of allowed and forbidden trajectories but a region of chaotic behaviour.
 This is especially true at low rigidities where quick oscillations can be observed.
 However, the approximation of the Earth magnetic field by a simple dipole together with the limited numerical precision makes an interpretation of the significance of the quick alternating focusing coefficients difficult.
 We are much more confident discussing average values, which is in this special case the value of 2.
 Consequently, in this picture there is an increase of the cosmic ray flux on top of the atmosphere by a factor of 2 over the rigidity range $[0-20~GV/c]$.
 
Considering the rigidity range between 3 and 12~GV/c, there are some clear physical structures, which are relatively narrow and which seem to follow a certain pattern (\autoref{80}).
 Above 12~GV/c, the structures almost completely disappear and the focusing coefficient converges to 1, as it is expected because there is only very little magnetic bending and therefore only very little focusing or dispersion at such high rigidities.
 This expected convergence towards 1 at high rigidities has been observed for all studied configurations.
 However, the rigidity where the rapidly alternating structures end and the convergence starts depends on latitude.

\begin{figure*}
\begin{minipage}{2\columnwidth}
\begin{center}
\includegraphics[width=0.95\columnwidth]{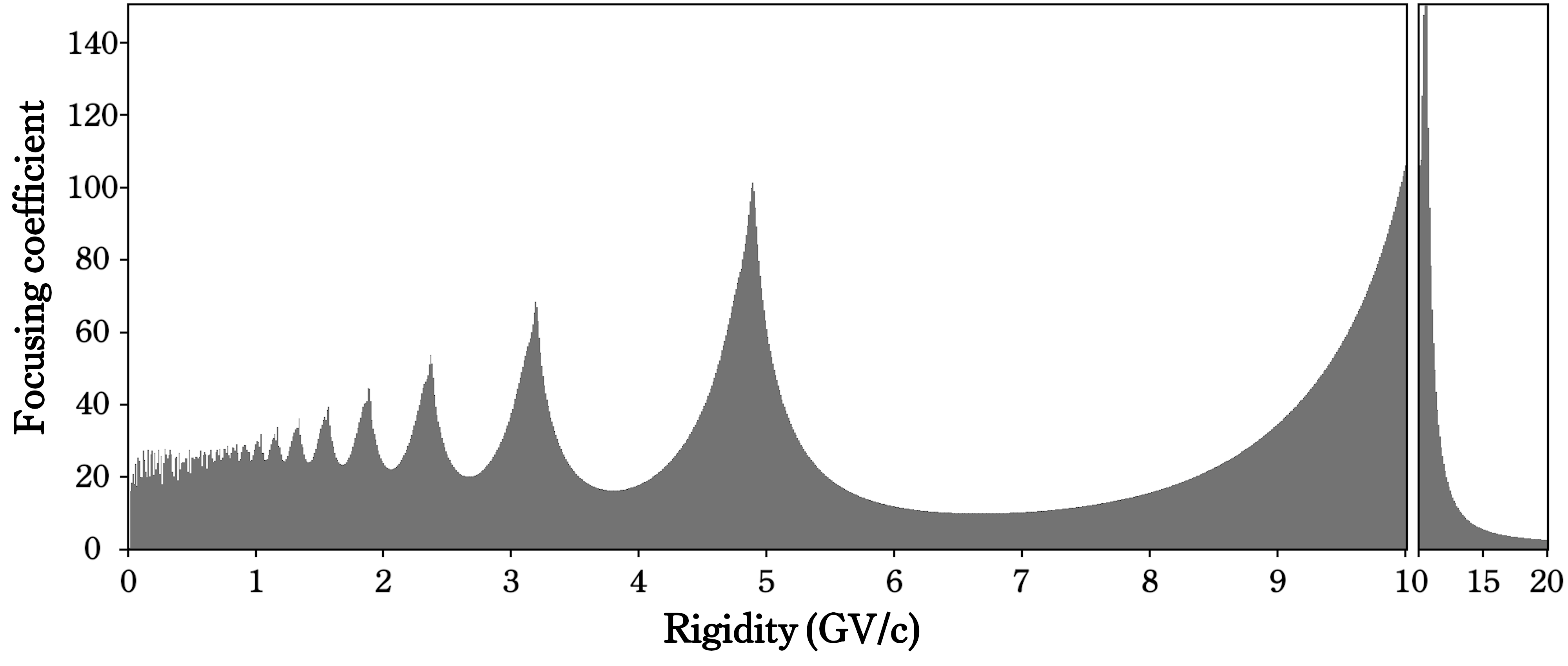}
\caption{\label{zen} Same as \autoref{80} but for a zenith angle of $0^\circ$.}
\end{center}
\end{minipage}
\end{figure*}

That the resonances are not numerical artefacts but are more general features can be seen in \autoref{zen}, where the parameters for the calculation have been the same as in \autoref{80} but this time for a zenith angle of $0^\circ$.
 At first glance, the resonances are very interesting structures but we don't consider them relevant for the calculation of galactic cosmic ray particle spectra, because, {\it first}, the structures are too narrow and, {\it second,} the occurrence and the size of the resonances are highly dependent on the structure of the Earth magnetic field.
 In this respect, the geocentric dipole approximation is likely not precise enough for a reliable interpretation of such resonance structures and further studies with more realistic magnetic field configurations are needed.

A second, and likely even more relevant change caused by the funnel effect is for the amplitude of the focussing coefficient and therefore for the flux densities of the determined cosmic ray particle spectra.
 Remember, without considering the funnel effect the focusing coefficient is either 0 for forbidden trajectories or 1 for allowed trajectories.
 Consequently, the average value will be below 1.
 In contrast, by considering the funnel effect, the focusing coefficient can reach very large values; in principle the value can be in the range from 0 to $+\infty$.
 For most of the studied cases the trajectories are either globally focused or dispersed for a large range of rigidities (for a given latitude).
 For the example shown in \autoref{80}, essentially all trajectories are allowed and the average of the focusing coefficient is in the range of 2.
 Consequently, the cosmic ray spectrum calculated for the top of the atmosphere by considering the funnel effect would be a factor of $\sim$2 higher than the spectrum calculated without considering the funnel effect.
 This, indeed, will have some consequences for cosmogenic nuclide production in the Earth atmosphere.

By considering the funnel effect we calculate a global decrease of the cosmic ray flux for latitudes below 40$^\circ$ and a global increase in the cosmic ray flux above 40$^\circ$.
 The total cosmic ray particle flux reaching the top of the atmosphere integrated over all latitudes is slightly increased.
 However, the significance of this finding is difficult to judge because it is based on the approximation that the Earth magnetic field is a simple geocentric dipole.

It is obvious that both effects found, the resonance-type structures and the high average focusing coefficient, are of importance because they both have the ability to change the shape and the magnitude of the cosmic ray particle spectrum on top of the atmosphere and therefore might affect cosmogenic nuclide production in the Earth atmosphere.
 However, both effects might also be of importance for studies of the cosmic ray particle spectrum outside the geomagnetic field.
 We will briefly elaborate on this topic in the next chapter.

\section{consequences for GCR spectra calculations}
\label{consequences}

The two effects discussed before, the occurrence of resonances and the increase of the total cosmic ray spectrum on top of the atmosphere, can also be of importance for studies of the galactic cosmic ray particle spectrum itself.
 We demonstrate this using a concrete example.
 Doing so, we focus on the BESS balloon experiment \cite{bess}, which is well suited for our purpose because the latitude of the flight was relatively constant (in the coordinate system of the magnetic field), which makes the computation of the funnel effect easier.
 The flight has been carried out in the latitude range $[60^\circ-80^\circ]$ and in the range of zenith angles $[0^\circ-35^\circ]$ (all in the coordinate system of the magnetic field).
 We computed the funnel effect and calculated averages for the focusing coefficients.
 Since we expect that some (or all) of the resonances disappear by using a more realistic version of the Earth magnetic field and especially by considering time and space variations (see below), we ignore the resonances and produced a smooth analytical function describing the baseline of the focussing coefficients, i.e., a baseline cutting essentially all resonances.
 Using such a smoothed version for the focussing coefficient avoids over-interpreting the results.

\begin{figure*}
\begin{minipage}{2\columnwidth}
\begin{center}
\includegraphics[width=\columnwidth]{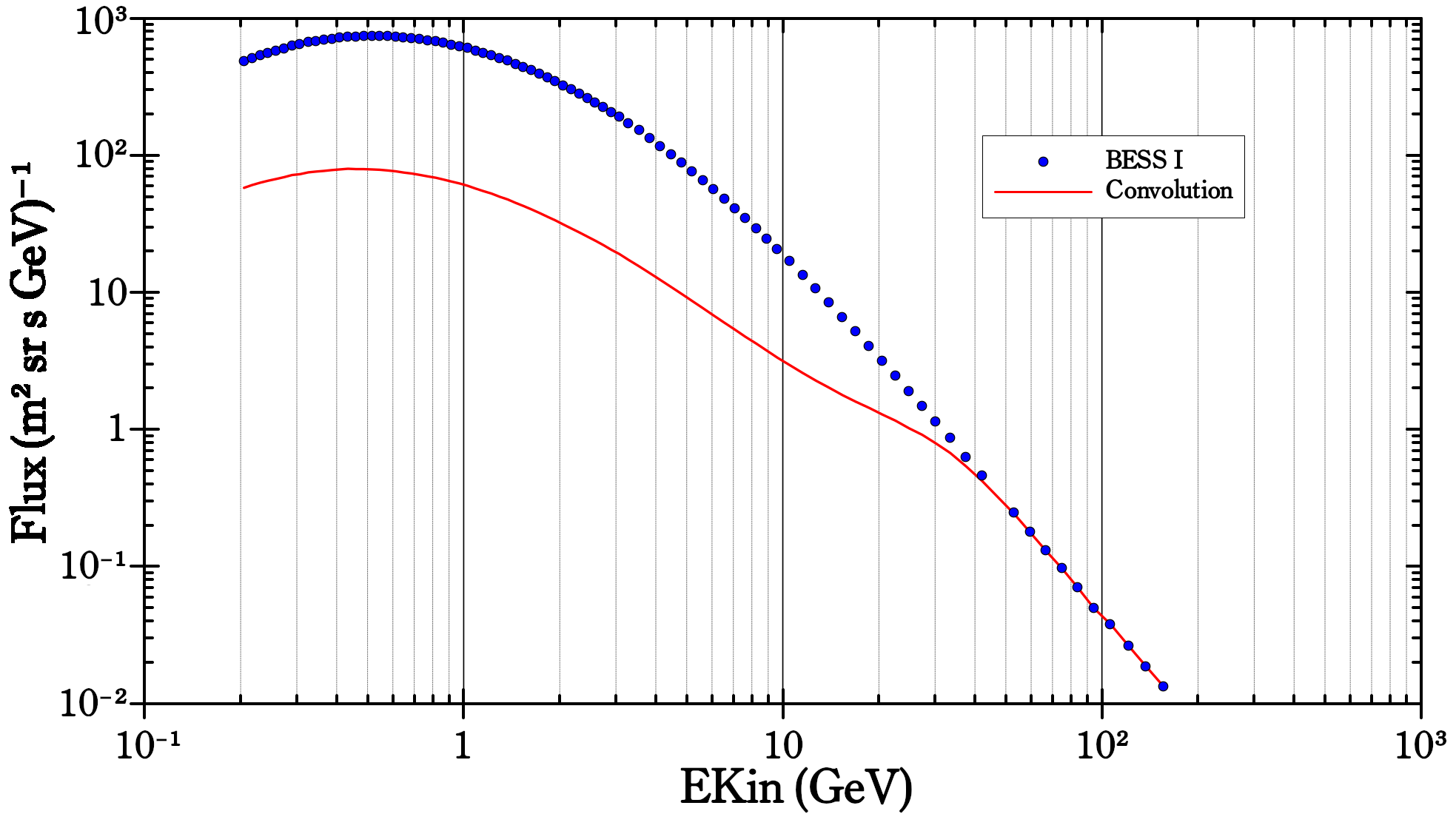}
\caption{\label{convo} Cosmic ray proton spectrum in the energy range 100~MeV-200~GeV.
 The blue bullets show the results given by the BESS collaboration calculated without considering the funnel effect.
 The red line is the result obtained by us calculated by fully considering the funnel effect (but by using a simplified magnetic field configuration)}
\end{center}
\end{minipage}
\end{figure*}

\autoref{convo} depicts our results for the GCR proton spectrum outside the magnetosphere calculated based on the experimental BESS-I data \cite{bess} (red line).
 The original result given by the BESS-collaboration, which has been calculated using the standard approach without considering the funnel effect, is shown by the blue dotted line.
 By considering the funnel effect, the total GCR flux outside the geomagnetic field is calculated lower, and for some energies significantly lower, than the spectrum calculated without considering the funnel effect.
 For the lowest considered energies, the flux calculated by us is about one order of magnitude lower than the BESS-result.
 The reduction in intensity holds up to energies in the range $\sim$30-40 GeV, where it is still in the range 10-40\%, i.e., the reduction is still significant.

Here we must emphasize that the deconvolution of the BESS-I data done by us is clearly an oversimplification and more realistic and reliable approaches must based on more realistic magnetic field configurations.
 However, the results clearly demonstrate that the funnel effect must be considered for a high quality deconvolution of the cosmic ray spectrum measured in the geomagnetic field for obtaining the original spectrum outside the magnetosphere.

\section{Conclusion}
\label{conc}

This study aims at demonstrating the importance of the funnel effect, i.e., the focusing and dispersion of cosmic rays in the magnetosphere.
 This effect, which has the ability to significantly modify the amplitude and the shape of the irradiation spectrum, is important in two directions.
 {\it First}, starting with the (original) GCR spectrum outside the Earth magnetic field, the funnel effect changes the amplitude and the shape of the cosmic ray particle spectrum on top of the atmosphere.
 This has some consequences for nuclide production in the Earth atmospheres and in the Earth surface.
 Since the funnel effect is less pronounced for high energetic particles, which are the dominant source particles for cosmogenic nuclide production in the Earth surface (via the production of secondary neutrons and muons), the changes for the terrestrial cosmogenic production rates are expected to be only small to moderate.
 However, the so-called scaling factors, i.e., the dependence of the terrestrial nuclide production on geomagnetic latitude and altitude needs to be carefully re-studied in view of our new results by considering the funnel effect in a realistic magnetic field configuration.

{\it Second}, the funnel effect also has the ability to affect deconvolution procedures needed to determine the cosmic ray particle spectrum outside the geomagnetic field from data measured inside the geomagnetic field.
 In this paper we studied the BESS balloon data as an example.
 Starting with the data measured inside the Earth geomagnetic field, we calculated, by considering the funnel effect, a cosmic ray particle spectrum outside the geomagnetic field that is up to one order of magnitude lower than the particle spectrum calculated without considering the funnel effect.
 The differences depend on the energy; they are largest at low energies and they disappear above $\sim$40 GeV.
 It is therefore safe to conclude that the process of dispersion and focusing of cosmic ray trajectories in the magnetosphere can have a major impact on the cosmic ray particle spectrum, especially at high latitudes.
 This calls for a revisit of the data obtained in experiments that have been performed deep inside the magnetosphere, like BESS \cite{bess}, AMS \cite{ams}, and PAMELA \cite{pamela}.

It is important to clearly mention the limitations of this analysis.
 For the current study we assume that the Earth magnetic field is a perfect dipole, which is not true; the magnetosphere is much more complex than this.
 For example, the solar wind induces strong asymmetries and significant time variations.
 This oversimplification clearly limits the  predictive power of our analysis and, therefore, the results are not considered quantitative accurate.
 However, we consider the major finding, that the funnel effect must be considered for high quality cosmic ray studies, as robust and relevant enough to warrant further studies using more realistic and possibly even dynamic magnetic field configurations.

\section*{Acknowledgements}

This work has been supported by the Swiss National Science Foundation (SNF 200021-159562 and 200020-182447).


\begin{thebibliography}{99}
\addcontentsline{toc}{chapter}{\hspace{5mm}Bibliography}

\bibitem{DavidLeya2019}
        David J.-C., and Leya I. 2019. \emph{``Spallation, cosmic rays, meteorites, and planetology.''}
        \href{https://doi.org/10.1016/j.ppnp.2019.103711}{Progress in Particle and Nuclear Physics, 109, id. 103711}

\bibitem{Dunai2010} 
		Dunai T. 2010. \emph{``Cosmogenic Nuclides: Principles, Concepts and Applications in the Earth Surface 
		Sciences.''}
        \href{https://doi.org/10.1017/CBO9780511804519}{DOI:10.1017/CBO9780511804519 Cambridge University Press.}

\bibitem{Ammonetal2008}
		Ammon K., Masarik J., Leya I. 2008. \emph{``New model calculations for the production rates of
		cosmogenic nuclides in iron meteorites.''}
        \href{https://doi.org/10.1111/j.1945-5100.2009.tb00746.x}{Meteoritics \& Planetary Science 44, 485-503}

\bibitem{LeyaMasarik2009}
		Leya I. and Masarik. 2009. \emph{``Cosmogenic nuclides in stony meteorites revisited.''}
        \href{https://doi.org/10.1111/j.1945-5100.2009.tb00788.x}{Meteoritics \& Planetary Science 44, 1061-1086}
        
\bibitem{Lietal2017}
		Li Y., Zhang X., Dong W., Ren Z., Dong T., Xu A. 2017. \emph{``Simulation of the production rates
		of cosmogenic nuclides on the Moon based on Geant4.''}
        \href{https://doi.org/10.1002/2016JA023308}{Journal of Geophysical Research: Space Physics 122, 1473-1486}

\bibitem{Argentoetal2015a}
		Argento D.C., Stone J.O., Reedy R.C., O'Brian K. 2015a. \emph{``Physics-based modeling of
		cosmogenic nuclides part I - Radiation transport methods and new insights.''}
        \href{https://doi.org/10.1016/j.quageo.2014.09.004}{Quaternary Geochronology 26, 29-43}
		
\bibitem{Argentoetal2015b}
		Argento D.C., Stone J.O., Reedy R.C., O'Brian K. 2015b. \emph{``Physics-based modeling of
		cosmogenic nuclides part II - Key aspects of in-situ cosmogenic nuclide production.''}
        \href{https://doi.org/10.1016/j.quageo.2014.09.005}{Quaternary Geochronology 26, 45-55}

\bibitem{MasarikBeer1999}
		Masarik J. and Beer J. 1999. \emph{``Simulation of particle fluxes and cosmogenic nuclide
		production in the Earths atmosphere.''}
        \href{https://doi.org/10.1029/1998JD200091}{Journal of Geophysical Research 114, 10299-12112}
		        
\bibitem{bess}
        BESS-Polar Collaboration.
		\emph{``Measurements of Cosmic-Ray Proton and Helium Spectra from the BESS-Polar Long-Duration Balloon Flights over Antarctica''}
		\href{http://dx.doi.org/10.3847/0004-637X/822/2/65}{Astrophys. J. 822, 65 (2016)}
		
\bibitem{ams}
        AMS Collaboration.
        \emph{``Precision Measurement of the Proton Flux in Primary Cosmic Rays from Rigidity 1 GV to 1.8 TV with the Alpha Magnetic Spectrometer on the International Space Station ''}.
        \href{https://doi.org/10.1103/PhysRevLett.114.171103}{PRL 114, 171103 (2015)}
		
\bibitem{pamela}
        N. Marcelli et al.
        \emph{``Time Dependence of the Flux of Helium Nuclei in Cosmic Rays Measured by the PAMELA Experiment between 2006 July and 2009 December ''}.
        \href{https://doi.org/10.3847/1538-4357/ab80c2}{Astrophys. J. 893, 145 (2020)}

\bibitem{Schereretal2006}
		Scherer K., Fichtner H., Borrmann T., Beer J., Desorgher L., Fl\"uckiger, Fahr H.-J., 
		Ferreira S.E.S., Langner U.W., Potgieter M.S., Heber B., Masarik J., Shaviv N.J., 
		Veizer J. \emph{``Interstellar-terrestrial relations: variable cosmic environments, the dynamic
		heliosphere, and their imprints on terrestrial archives and climate''}
        \href{https://doi.org/10.1007/s11214-006-9126-6}{Space Science Reviews, 127, 327-465}  

\bibitem{Stoermer1950}
		St\"ormer C. 1950.
		\emph{``The polar aurora.''}
		\href{https://doi.org/10.1002/qj.49708235123}{Oxford University Press, London.}

\bibitem{Cooketal1991}
		Cooke D.J., Humble J.E., Shea M.A., et al. 1991. 
        \emph{``On cosmic-ray cut-off terminology''}.
		\href{https://doi.org/10.1007/BF02509357}{Il Nuovo Cimento C Phys. C 14, 213.}
		
\bibitem{Smart2000}
        D.F. Smart, M.A. Shea, and E.O. Flückiger.
        \emph{``Magnetospheric Models and Trajectory Computations''}.
        \href{https://doi.org/10.1023/A:1026556831199}{Space Sci. Rev. 93 (1-2), 305 (2000)}

\bibitem{Kasper1959}
		Kasper J.E.
		\emph{``The Earths simple shadow effect on cosmic radiation''} 
		\href{https://doi.org/10.1007/BF02724760}{Nuovo Cimento, 11, 1 (1959)}


\end{thebibliography}
\end{document}